\documentclass[letterpaper,12pt,english]{article}

\usepackage{amsfonts,bm,amssymb,euscript,array,babel,cite}
\usepackage{amsmath,amsthm} 
\usepackage[dvips]{epsfig}



\newcommand{\be}{\begin{equation}} \newcommand{\ee}{\end{equation}}
\newcommand{\beq}{\begin{equation}} \newcommand{\eeq}{\end{equation}}
\newcommand{\beqa}{\begin{eqnarray}}
\newcommand{\eeqa}{\end{eqnarray}} \newcommand{\eq}[1]{(\ref{#1})}
\newcommand{\cg}[7]{
 #1 \!\!\!
 {\scriptsize 
 \begin{array} {ccc}
#2 & \!\!\!\!\!\!\!#3 &\!\!\!\!\!\!\!#4 \\
#5 & \!\!\!\!\!\!\!#6 &\!\!\!\!\!\!\!#7
\end{array}}} 
\def\nn{\nonumber} \def\bea{\begin{eqnarray}} \def\eea{\end{eqnarray}}
\def\obar{\overline}

\def\a{\alpha}  
  
 \def\d{\delta} 
  
    \def\k{\kappa}
 \def\L{\Lambda}  
    \def\r{\rho}
\def\s{\sigma}   \def\th{\theta}

\def\cA{{\cal A}}  \def\cC{{\cal C}}    
\def\cH{{\cal H}}     \def\cM{{\cal M}} 
\def\cO{{\cal O}}  \def\cQ{{\cal Q}}

\def\R{{\mathbb R}} \def\C{{\mathbb C}} \def\N{{\mathbb N}}
\def\Z{{\mathbb Z}} \def\one{\mbox{1 \kern-.59em {\rm l}}}

\def\bit{\begin{itemize}} \def\eit{\end{itemize}} \def\Tr{\mbox{Tr}}

\def\({\left(} \def\){\right)}

\def\del{\partial}
\def\Ric{{\rm Ric}}

\renewcommand{\Tr}{{\rm Tr}}

 \allowdisplaybreaks[3]

\begin{document}

\renewcommand{\title}[1]{\vspace{10mm}\noindent{\Large{\bf
#1}}\vspace{8mm}} \newcommand{\authors}[1]{\noindent{\large
#1}\vspace{5mm}} \newcommand{\address}[1]{{\itshape #1\vspace{2mm}}}

\begin{titlepage}
\begin{flushright}
UWThPh-2013-22\\
\end{flushright}

\begin{center}

\title{ \Large 2D fuzzy Anti-de Sitter space from matrix models}

\vskip 3mm

 \authors{Danijel Jurman${}^{* ,}${\footnote{Danijel.Jurman@irb.hr }}, 
 Harold Steinacker${}^{\dagger ,}${\footnote{harold.steinacker@univie.ac.at}}
 }
 
\vskip 3mm

 \address{ 
  ${}^*$ {\it 
  Theoretical Physics Division, Rudjer Boskovic ́Institute \\ 
   P.O. Box 180, 10002 Zagreb, Croatia } \\[3ex]
  
 ${}^{\dagger}$  {\it Faculty of Physics, University of Vienna \\ 
 Boltzmanngasse 5, A-1090 Vienna, Austria  }  
 }

\vskip 1.4cm

\textbf{Abstract}

\vskip 3mm

\begin{minipage}{14cm}

We study the fuzzy hyperboloids $AdS^2$ and $dS^2$ as brane solutions in matrix models. 
The unitary representations of $SO(2,1)$ required for 
quantum field theory are identified, and explicit formulae for their realization 
in terms of fuzzy wavefunctions are given. 
In a second part, we study the $(A)dS^2$ brane geometry and its dynamics, as
governed by a suitable matrix model. 
In particular, we show that trace of the 
energy--momentum tensor of matter induces transversal perturbations of the brane and of the Ricci scalar.
This leads to a linearized form of Henneaux--Teitelboim--type gravity, illustrating
the mechanism of emergent gravity in matrix models.

\end{minipage}

\end{center}

\end{titlepage}

 \tableofcontents
\section{Introduction}

There has been a great amount of work on noncommutative field theory on the the fuzzy sphere and similar
compact quantum spaces. 
Part of their appeal stems from the fact that the space of functions on these spaces 
has a simple group-theoretical structure and is finite-dimensional,
reflecting their finite symplectic volume. This leads to mathematically well-controlled toy models for 
 noncommutative field theory and geometry, see e.g. 
 \cite{Madore:1991bw,Grosse:1995ar,Chu:2001xi,Vaidya:2001rf,Karabali:2001te,Steinacker:2003sd,
 CastroVillarreal:2004vh,Balachandran:2005ew,DelgadilloBlando:2008vi} 
and references therein.
However, most of the work so far has been for spaces with Euclidean signature, and it would be desirable to 
know more about fuzzy spaces with Minkowski signature.

In this paper, we study in detail 2-dimensional fuzzy 
de Sitter space $dS^2$ and Anti-de Sitter space $AdS^2$, 
which are quantized homogeneous spaces with Minkowski signature and non-vanishing curvature. 
Fuzzy $AdS^2$ has been studied previously in \cite{Ho:2000fy,Ho:2000br}. 
In the first part of this paper, we elaborate the space of functions on these  fuzzy hyperboloids, 
and provide explicit formulae
for the square-integrable wavefunctions corresponding to unitary irreducible representations of $SO(2,1)$.
For the discrete series representations we recover previous results obtained in \cite{Ho:2000fy},
and for the principal continuous representations our results are new. 
This provides the basic constituents for quantum field theory\footnote{For 
a discussion in the undeformed case see e.g. \cite{Joung:2006gj} and references therein.}
 on fuzzy  $AdS^2$ and $dS^2$. 
In particular, this also allows to establish the required quantization map
for the fuzzy geometry.

In a second part, we consider a matrix model which describes dynamical  
fuzzy  $AdS^2$ and $dS^2$ spaces as brane solutions.
As discussed in \cite{Steinacker:2008ri}, this leads to a dynamical effective geometry 
on the branes, determined by a combination 
of the embedding geometry of the brane and its Poisson structure.
The present 2-dimensional example 
provides an interesting toy model for emergent gravity, with 
a non-trivial curvature background. We study the perturbations around the $AdS^2$ solutions,
and their dynamics in the presence of matter.
This is interesting because the extrinsic curvature of the brane leads to a 
coupling of the linearized matrix perturbations to the energy-momentum tensor\footnote{rather than just its 
derivative, as on trivially embedded branes.}, as 
 pointed out in \cite{Steinacker:2012ra,Steinacker:2009mp}.
 More precisely, the transversal perturbations of the brane couple to the trace of the 
energy-momentum tensor of matter, due to the extrinsic curvature. 
It turns out that the  perturbations of the effective metric are governed by 
a linearized Henneaux--Teitelboim--type gravity \cite{Brown:1986nm}, relating 
the trace of the energy-momentum tensor  to the  Ricci scalar.
This is remarkable, because it results directly from the underlying matrix model action,
without adding any gravity action.
It provides a simple example for the mechanism  of emergent gravity in Yang-Mills matrix models. 
However, this result is restricted to the linearized regime.

In 4 and higher dimensions, the dynamics of the effective geometry is complicated due to a mixing 
between tangential and transversal brane perturbations \cite{Steinacker:2012ra}, 
which prohibits a full  understanding at present.
A similar mixing is observed here, but we are able to disentangle the coupled wave equations, 
and thereby essentially solve the perturbative dynamics. 
Therefore the present 2-dimensional case should serve as a useful step towards understanding the more complicated  
higher-dimensional case.

\section{Classical two-dimensional hyperboloid}

\subsection{Geometry and isometry group}

There are three types of two-dimensional non-compact  spaces with constant curvature,
given by the Anti-de Sitter space $AdS^2$, de Sitter space $dS^2$  and the 
hyperbolic or Lobachevsky plane  $H^2$. In this paper we discuss $AdS^2$ and $dS^2$, which 
can be naturally realized  as the one-sheeted hyperboloid embedded in $\R^3$  
through
\be
x^a x^b\, \eta_{ab} = 
-(x^1)^2 - (x^2)^2 + (x^3)^2 = -R^2.
\label{class-quadrik}
\ee
In terms of conformal coordinates $-\pi/2 < \sigma < \pi/2$ and $\pi<\tau\leq \pi$,
the embedding of classical Anti-de Sitter space $AdS^2$ is given by   
\bea
x^1=R {\cos \tau \over \cos \sigma},\;x^2=R {\sin\tau \over \cos \sigma}
,\;x^3=R \tan \sigma.
\label{glcoord}
\eea
The induced metric is pseudo-Riemannian 
\bea
 &&g_{\mu\nu} = \eta_{ab}\partial_\mu x^a \partial_\nu x^b,\;\eta_{ab} = {\rm diag}(-1,-1,1),\;\mu,\nu=\sigma,\tau\;,\\
 &&g_{\tau \tau}= {R^2 \over \cos^2 \sigma},\;g_{\sigma \sigma}=-{R^2 \over \cos^2 \sigma},\;g_{\sigma \tau}=0,
\label{indmet}
 \eea
with closed time-like circles\footnote{They can be avoided by
passing to the universal cover of $AdS^2$.} around $x^3 =$ const . 
De Sitter space $dS^2$ is obtained from $AdS^2$ by switching the roles 
of the time and space, thus changing the overall sign in the metric. 
The circles $x^3=$ const are then space-like, and there are no closed time-like curves.

Both $AdS^2$ and $dS^2$ admit the group $SO(2,1)$ or its cover $SU(1,1)$ as isometries,
generated  by vector fields $K^a, \ a = 1,2,3$ 
\be
K^1\!=-i\cos \!\tau \sin\! \sigma \partial_\tau - i\sin\!\tau \cos\! \sigma \partial_\sigma,\;
K^2\!=-i\sin\! \tau \sin \!\sigma \partial_\tau+i\cos \!\tau \cos\! \sigma \partial_\sigma,\;K^3\!=-i\partial_\tau, 
\label{vecfieldalg}
\ee 
which close $su(1,1)$ Lie algebra with respect to commutators
\be
[K^a,K^b] = i f^{ab}_{\ \, c}\, 
K^c
\ee
or explicitly
\bea
[K^1,K^2] &=& -i K^3,\;[K^2,K^3] = i K^1,\;[K^3,K^1] = i K^2.
\label{su11-algebra}
\eea
The Casimir operator of $su(1,1)$ Lie algebra is defined as
\be
C = - ({K^1})^2 - ({K^2})^2 + ({K^3})^2 .
\ee
As usual, it is convenient to introduce the ladder operators
\be
K^\pm=K^1\pm i K^2,
\label{ladderop}
\ee
which satisfy the commutation relations 
\be
\left[ K^3, K^\pm \right] =\pm K^\pm,\;\left[ K^+,K^-\right]=-2K^3.
\label{kpmalg}
\ee
Then unitary irreducible representations of $SO(2,1)$ are spanned by a basis
$|j,m\rangle$ of weight states, where $j$ is related to the eigenvalue 
of the Casimir $C$, and $m$  is the eigenvalue of $K^3$ and 
the action of $K^\pm$ on $|j,m\rangle$ produces a state with weight $m\pm1$:
\be
K^3 K^\pm |j,m\rangle =(m\pm 1) K^\pm |j,m\rangle \sim |j,m\pm 1\rangle .
\label{chainsim}
\ee
A chain of states obtained by
the successive action of $K^-$ operator terminates if there exist state such that 
\be 
K^- |j,m_0\rangle =0.
\label{minweight}
\ee
Denoting this lowest weight by $j=m_0$, it follows that
\be
0=K^+K^- |j,j\rangle =\left(-C+K^3(K^3-1)\right) |j,j\rangle \Rightarrow C=j(j-1)
\label{jeqm0}
\ee
Therefore the chain of states which span this  
irreducible lowest weight representation is determined by the state $  |j,j\rangle$ of lowest weight, via
\be
 |j,j+m\rangle \sim {K^+}^m  |j,j\rangle .
\label{dislowchain}
\ee
By analogy, the highest weight representation are obtained by interchanging roles of $K^+$ and $K^-$ operators.
If no lowest or highest weight state  exists, then the normalisability condition implies
$C<0$, and the states belong to the unitary irreducible continuous representations.

In general,
the resulting structure of irreducible representations is as follows:
\bea
K^3|j,m\rangle&=&m|j,m\rangle, \nn\\
K^+|j,m\rangle&=&a_{m+1}|j,m+1\rangle, \nn\\
K^-|j,m\rangle&=&a_m|j,m-1\rangle,
\label{expabsalg}
\eea
where
\be
a_m=\sqrt{m(m-1)- j(j-1)}.
\label{norma}
\ee
The finite-dimensional irreducible representations of $SU(1,1)$ are obtained for 
$j  \in - \N/2$. They are not unitary, and correspond to the spin $|j|$ 
representations $V_{|j|}$ of $SU(2)$ with $C = -|j|(|j|+1)$.
All unitary irreducible representations are infinite-dimensional, 
and fall into one of the following 
classes\footnote{We only consider representations with integer weights for simplicity.} \cite{Bargmann:1946me}:
 
$\bullet$ The discrete series of the highest and the lowest weight representations 
\bea
D_j^{+}, \quad j \in \N_{>0}: \qquad  \cH_j &=& \{|j,m\rangle; m = j,j+1,\cdots ;m \in \N\; \},  \nn\\
D_j^-, \quad j \in \N_{>0}: \qquad \cH_j  &=& \{|j,m\rangle; m = -j,-j-1,\cdots ;-m \in \N\;\},
\label{discrete}
\eea
characterized by $C=j(j-1)\geq 0$.

$\bullet$ The principal continuous series 
\be
P_s, \quad s \in \R,\quad 0<s<\infty,\quad  j = \frac 12 + i s, 
\qquad \cH_j = \{|j,m\rangle; m=0,\pm 1, ...; m  \in \Z \}
\label{principal}
\ee
labeled by a real number $s$ and $C= -\left( s^2 + \frac 14\right)<-1/4$.

$\bullet$ The complementary series
\be
P_j^c, \quad  1/2 < j < 1,\qquad j \in \R ,
\qquad \cH_j = \{|j,m\rangle; m=0,\pm 1, ...; m  \in \Z \}
\label{complemenary}
\ee
with $-1/4<C<0$.

\subsection{Functions and Poisson bracket}

In order to carry out the quantization of  $(A)dS^2$, it is useful to organize the space of functions on $(A)dS^2$
in terms of irreducible representations of $SU(1,1)$. This provides at the same time the basis of 
eigenfunctions of the invariant d'Alembertian $\Box_g$, 
\be 
\Box_g = {1\over \sqrt{|g|}} \partial_\mu \sqrt{|g|} g^{\mu \nu} \partial_\nu=
{\cos^2 \sigma \over {R^2}}\left( \partial^2_\tau-\partial^2_\sigma\right)=
{{K^1}^2+{K^2}^2-{K^3}^2 \over {R^2}},
\label{box}
\ee
which is related to the Casimir operator of   $su(1,1)$.
Here $g=\det(g_{\mu\nu})$,  and $g^{\mu\nu}$ is inverse of the metric.
We can thus decompose any  function on the hyperboloid into eigenfunctions of $\Box_g$,
\be
\Box_g \phi + \alpha \phi=0\ ,
\label{scfieldeq}
\ee 
and label the solutions by $j$ and $m$ as above.
The solutions corresponding to the finite-dimensional  representations 
are realized by polynomial functions $\rm{Pol}(x^a)$; they are of course not normalizable on $(A)dS^2$.
The square-integrable functions corresponding to unitary irreducible representations are
given explicitly  in terms of hyper-geometric functions
\be\label{solcl}
\phi_{jm}\!=\!e^{-im \tau}\!\!
\cos^j \!\sigma\!\left[ \!a\; _2F_1\!\!
\left(\!\!{j\!+\!m \over 2},\! {j\!-\!m \over 2},\!{1\over 2};\!\sin^2\!\sigma\!\right)\!+\!
b \sin\!\sigma \;_2F_1\!\!
\left(\!\!{j\!+\!m\!+\!1 \over 2},\! {j\!-\!m\!+\!1 \over 2},\!{3\over 2};\!\sin^2\!\sigma\!\right)\!\!\right],
\ee
where 
\be
C = j(j-1)=R^2\alpha
\ee
is the Casimir.
For  $AdS^2$, the scalar fields  corresponding to positive or negative energy unitary representations
belong to the discrete representation $D_j^\pm$, with $\a>0$. 
Then the equations for the lowest weight state have a unique solution 
\be\label{unisol}\left.
\begin{array} {c}
K^3 \phi_{jj}=j\phi_{jj}\\
K^- \phi_{jj}=0
\end{array}\right\}\Rightarrow \phi_{jj}\sim e^{ij\tau} \cos^j \sigma 
\ee
and the spectrum is non-degenerate. 
On the other hand the states given by (\ref{solcl}) with $\a<0$
belong to the continuous representations, with two-fold degenerate spectrum. These are the physical scalar fields on 
de Sitter space $dS^2$.
Putting these together, we have the following decomposition of functions on the hyperboloid $(A)dS^2$
\begin{align}
 L^2((A)dS^2) = \oplus_{J \geq 1} D_J^+ \oplus_{J \geq 1} D_J^- \oplus 2\int_0^\infty dS P_S
\label{functions-decomp}
 \end{align}
along with the space of polynomial functions $\rm{Pol}(x^a)$.

In the following we discuss fuzzy versions of these non-compact spaces, and their associated spaces of functions.
As a starting point, we note that 
the natural $SO(2,1)$-invariant volume element endows the hyperboloid with a non-degenerate symplectic form
\be
\omega=\frac{R}{\kappa \cos^2 \sigma } d\tau \wedge d \sigma
\label{volel}
\ee 
with $d\omega = 0$, introducing a scale perameter $\k$. 
Its inverse defines the Poisson bracket of two functions
\be 
\{ f,g \}=  {\kappa \cos^2 \sigma \over  R} \left( \partial_\tau f \partial_\sigma g-
\partial_\sigma f \partial_\tau g\right). 
\label{Poissonbra}
\ee
We can now look for a quantization of this Poisson manifold $\cM$, cf. \cite{Bordemann:1993zv}.
This means that the algebra of functions $\cC(\cM)$ should be mapped to 
a non-commutative (operator) algebra $\cA$, such that the commutator is approximated
by the Poisson bracket. In the present case, 
the group-theoretical structure of $(A)dS^2$ provides a natural and explicit quantization,
in analogy to the case of the fuzzy sphere \cite{Madore:1991bw}.
As a first step, we note that the Poisson brackets of the 
embedding functions $x^a$  satisfy the Lie algebra of $SO(2,1)$ 
\be 
\{x^a,x^b\} ={\kappa} f^{ab}_{\ \, c}\, x^c 
\label{cooralg}
\ee
where $f^{ab}_c$ are structure constants of $SO(2,1)$.
This implies as usual that the corresponding Hamiltonian vector fields satisfy the same Lie algebra, and 
indeed it is easy to verify that the $SO(2,1)$ vector fields \eq{vecfieldalg} are given by
\be
K^a  = \frac{i}{\kappa} \{x^a,.\} \ .
\ee

\section{Fuzzy hyperboloid}

In analogy to the fuzzy sphere \cite{Madore:1991bw},
we define fuzzy two-dimensional hyperboloid in terms of three hermitian matrices 
(or operators) $X^a$, 
which are interpreted as quantization of the embedding functions $x^a$.
In view of \eq{cooralg}, we impose the following relations
\bea
[X^a,X^b] &=& i \kappa\,f^{ab}_{\ \, c}\, X^c, 
\label{algemcoor}
\eea
where $f^{ab}_c$ are structure constants of the Lie algebra $su(1,1)$.
Therefore the $X^a$ are  rescaled $su(1,1)$
generators, and we assume that they act on a certain irreducible unitary
representation $\cH_j$ of the Lie algebra. 
We can then write the Casimir operator as 
\bea 
X^a X^b\,\eta_{ab} = \kappa^2 j(j-1) .
\label{Casimir-constraint} 
\eea
Since $\cH_j$ is assumed to be irreducible, the 
$X^a$  generate the full algebra  $\cA$ of operators on $\cH_j$ 
\be
\cA := End(\cH_j) \cong \cH_j \otimes \cH_j^*,
\label{matalg}
\ee
where $\cH_j^*$ is dual representation of $\cH_j$. This algebra is an infinite-dimensional vector space, which
naturally carries an action of $su(1,1)$ by conjugation with the generators $X^a$:
\be
K^a\triangleright \Phi = {1 \over \kappa} [X^a,\Phi], \qquad \Phi \in \cA .
\label{natact}
\ee
We  now specify the representation $\cH_j$. 
Since the matrices $X^a$ should be interpreted 
as quantized embedding functions $x^a$ of the hyperboloid and comparing the spectrum of $X^3$ with the range of 
$x^3 \in \langle -\infty,\infty\rangle$, 
we choose  $\cH_j$ to be a principal continuous 
representation\footnote{The complementary representation 
is rejected because it does not admit a semi-classical limit
for fixed curvature, as explained in section \ref{sec:semiclass}.}, 
in accord with \cite{Ho:2000fy}.

We can furthermore define an invariant scalar product
\bea\label{scalar product}
(\Phi_1,\Phi_2)=\Tr \Phi_1^\dagger \Phi_2,\qquad \Phi_1,\Phi_2 \in \cA.
\eea
$\cA$ contains in particular the polynomials generated by the $X^a$,
where this trace diverges.
However,  $\cA$ also contains normalizable matrices corresponding to physical scalar fields, which are of main interest here.
Finding such normalizable matrices is equivalent to decomposing $\cA=\cH_j \otimes \cH_j^*$ into irreducible
unitary representations of $su(1,1)$. This problem
has been extensively studied in the literature \cite{Repka,HolmanBiedenharn}.
In general, the states $|JM\rangle$ which belong to a particular unitary irreducible representation  in 
 $\cH_{j_1} \otimes \cH_{j_2}$ are given by 
\be
|JM\rangle=\sum_{m_1,m_2}
\cg{C}{j_1}{j_2}{J}{m_1}{m_2}{M}
|j_1 m_1 \rangle \otimes | j_2 m_2\rangle \ .
\label{genwig}
\ee
Here the $C$'s are the Wigner coefficients, which vanish unless $M=m_1+m_2$. In the special case of 
$\cH_j \otimes \cH_j^*$, we represent the state (\ref{genwig}) as a matrix $ \Phi^J_M$
\be
\label{CGcoeff}
\Phi^J_M=\sum_{m_1 m_2}\; 
\cg{D}{j}{j}{J}{m_1}{m_2}{M}
|j m_1 \rangle \langle j m_2|,
\ee
where the $D$'s vanish unless $M=m_1-m_2 \ \in \Z$. 
Since we chose the principal continuous representation 
$\cH_j \cong P_s$, 
one obtains the following decomposition of the space of functions $\cA$ into unitary modes \cite{Repka}: 
\be\label{decompo}
P_s \otimes P_s^\prime = \oplus_{J \geq 1} D_J^+ \oplus_{J \geq 1} D_J^- \oplus 2\int_0^\infty dS P_S \ ,
\ee
along with the space of polynomial functions $\rm{Pol}(X^a)$. In the next section 
we will recover this 
result and obtain the corresponding  fuzzy wavefunctions explicitly, which solve the eigenvalue equations
\bea
&& \eta_{ab} K^a\triangleright K^b\triangleright \Phi^J_M=-\frac 1{\kappa^2}\Box \, \Phi^J_M= {\eta_{ab}\over \kappa^2} [X^a,[X^b,\Phi^J_M]]=J(J-1)\Phi^J_M, \nn\\
&&K^3\triangleright \Phi^J_M={1\over \kappa} [X^3,\Phi^J_M]=M\Phi^J_M \ .
\label{matlap}
\eea

\subsection{Fuzzy wavefunctions}

We can determine the fuzzy wavefunctions $\Phi^J_M$ explicitly, using their definition
as irreducible representations of $SO(2,1)$.
As an element of the operator algebra $\cA$, the matrix $\Phi^J_M$
acts on  $|j n\rangle \in \cH_j$ as 
\be\label{actofJM}
\Phi^J_M |j n\rangle=
\sum_m \cg{D}{j}{j}{J}{m}{m\!-\!M}{M}
|j m \rangle \langle j m-M|j n\rangle=
\cg{D}{j}{\;j}{J}{n\!\!+\!\!M}{n}{M} |j n+M\rangle \ .
\ee
Defining the matrix $D^J_M(K_3)$ by its action on  $|jn\rangle$
\be\label{dis}
D^J_M(K_3)|jn\rangle = \cg{D}{j}{j}{J}{\;n}{n\!-\!M}{M}|jn\rangle
\ee
and using defining property $\Gamma(x+1)=x\Gamma(x) $ of Gamma function,
we can express $\Phi^J_M$ for integer $M$ using \eq{expabsalg} as 
\bea\label{formmat}
\Phi^J_M&=&D^J_M (K^3)\sqrt{\Gamma(K^3-M-j+1)\Gamma(K^3-M+j) \over \Gamma(K^3-j+1)\Gamma(K^3+j)}  
{K^+}^{M}=\nn\\
&=& D^J_M (K^3) \left({1 \over \sqrt{(K^3-j)(K^3+j-1)}}K^+\right)^M,\qquad M>0\label{phi1}\\
\Phi^J_M &=&{K^-}^M \sqrt{\Gamma(K^3-M-j+1)\Gamma(K^3-M+j) \over \Gamma(K^3-j+1)\Gamma(K^3+j)} D^J_M(K^3)=\nn\\
&=&  \left(K^-{1 \over \sqrt{(K^3-j)(K^3+j-1)}}\right)^M {D^J_M}(K^3)^\dagger ,\qquad M<0 \ .\label{phi2}
\eea
To derive the final expressions (\ref{phi1}) and (\ref{phi2}) one applies the identity 
\be\label{nekiid}
\sqrt{\Gamma(K^3-M-j+1)\Gamma(K^3-M+j) \over \Gamma(K^3-j+1)\Gamma(K^3+j)}  
{K^+}^{M}=\left({1\over \sqrt{(K^3-j)(K^3+j-1)}}K^+\right)^M,
\ee
which can be verified using 
\be\label{iden}
K^+F(K^3)=F(K^3-1)K^+ \ ,
\ee
 which follows from  (\ref{kpmalg}).
We note that the expressions (\ref{phi1}) and (\ref{phi2}) are  hermitian conjugates of each other. This reflects the 
fact that ${\Phi^J_M}^\dagger$ is a solution of (\ref{matlap}) with eigenvalue $-M$ if $\Phi^J_M$ is a solution with eigenvalue $M$.

The above considerations apply to any representation.
For the discrete series representations $D_J^+$  in (\ref{decompo}) with
$J$ being integer, the basis of states  is completely determined 
by the minimal weight state  annihilated by $K^-$. 
Acting with $K^-$ on (\ref{phi1}) we obtain  
\bea\label{minweigh}
&&[K^-,\Phi^J_M]=\sqrt{(M-J)(M+J-1)}\Phi^J_{M-1}=\nn\\
&&=\left[D^J_M(K^3+1)\sqrt{(K^3+j)(K^3-j+1)}-D^J_M(K^3)\sqrt{(K^3-M+j)(K^3-M-j+1)}\right]\times\nn\\
&&\times\left({1\over \sqrt{(K^3-j)(K^3+j-1)}}K^+\right)^{M-1}.
\eea
Specializing this to the case $M=J$, we see that the expression in square bracket must vanish
\bea\label{lowwghteq}
&&D^J_J(K^3+1)\sqrt{\Gamma(K^3+j+1)\Gamma(K^3-j+2)\over \Gamma(K^3-J-j+2)\Gamma(K^3-J+j+1)}-\nn\\
&&-D^J_J(K^3)\sqrt{\Gamma(K^3+j)\Gamma(K^3-j+1)\over \Gamma(K^3-J+j)\Gamma(K^3-J-j+1)}=0 \ .
\eea
Since $K^3$ takes only integer values here, we can conclude that 
\be\label{sollowwght}
D^J_J(K^3)=\sqrt{\Gamma(K^3-J+j)\Gamma(K^3-J-j+1) \over \Gamma(K^3+j)\Gamma(K^3-j+1)}
\ee
(up to  normalization).
Finally, for the lowest weight state in $D^+_J$ using (\ref{nekiid}) we obtain 
\be\label{lowestws}
\Phi^J_J=\left({1\over (K^3-j)(K^3+j-1)}K^+\right)^J
\ee
in agreement with findings of \cite{Ho:2000br,Ho:2000fy}.
The highest weight state in $D^-_J$ is given by hermitian conjugate of (\ref{lowestws}).

For the principal continuous representation $P_S$ with $J=1/2+iS$ in (\ref{decompo}), 
the matrices $D^J_M$ are solutions of second order difference equation obtained 
from (\ref{minweigh}) applying $K^+$ on it: 
\small
\bea 
&&\left[(M-J)(M+J-1)-(K^3-j)(K^3+j-1)-(M-K^3-j)(M-K^3+j-1)\right]C^J_M(K^3)=\nn\\
&&\sqrt{(K^3-j)(K^3+j-1)(M-K^3+j)(M-K^3-j+1)} C^J_M(K^3-1)+\nn\\
&&\sqrt{(K^3+j)(K^3-j+1)(M-K^3-j)(M-K^3+j-1)}C^J_M(K^3+1)  \label{secorddifeq}.
\eea
\normalsize
Here we find it convenient to write $D^J_M$ as 
\be\label{cgmat}
D^J_M(K^3)=C^J_M(K^3)e^{i\pi K^3},
\ee
where $C^J_M(K_3)$ is a matrix with elements given by the Wigner coefficients
for the principal continuous representations of $su(1,1)$ in \eq{decompo}.
This can be seen after noting that
this second order difference equation is a special case of equation for the general Wigner coefficients
as found in \cite{HolmanBiedenharn}.
Finally,  we can express $C^J_M$ in terms of two independent solutions\footnote{The solutions are degenerate in the case of $J$ being integer.}
\small
\be\label{solut}
C^J_M(K^3)=\sqrt{\Gamma(M-K^3-j+1)\Gamma(M-K^3+j) \over \Gamma(K^3+j)\Gamma(K^3-j+1)}{aG^J_M(K^3)+bG^{1-J}_M(K^3) \over
\Gamma(M-K^3+J-j+1)\Gamma(M-K^3-J-j-2)}.
\ee
\normalsize
where $G^J_M(K^3)$ is the hypergeometric series
\be\label{geis}
G^J_M(K^3)=_3\!\!F_2(J,2j+J-1,J-M;K^3+J+j-M,2J),
\ee
defined by 
\bea
_3F_2(a,b,c;d,e)=\sum_{k=0}^\infty {\Gamma(a+k)\Gamma(b+k)\Gamma(c+k)\Gamma(d)\Gamma(e) \over k!\Gamma(a)\Gamma(b)\Gamma(c)\Gamma(d+k)\Gamma(e+k)}.
\eea
To summarize, we have obtained explicit matrices $\Phi^J_M$ of the form
\begin{align}
 \label{matform}
\Phi^J_M =\left\{\begin{array}{rl}
  F^J_M(X^3){X^+}^M, & M\geq 0 \\
  \tilde F^J_M(X^3){X^-}^M, & M\leq 0 \\
\end{array} \right. ,
\end{align}
 realizing the 
decomposition \eq{decompo} of $\cA = \cH_j \otimes \cH_j^*$ into unitary representations of $SO(2,1)$.
They form an orthonormal basis for the inner product defined by the trace \eq{scalar product}.
This is in one-to-one
correspondence with the decomposition \eq{functions-decomp} of
classical functions 
\begin{align}
 \label{matform-class}
\phi^J_M =\left\{\begin{array}{rl}
  f^J_M(x^3){x^+}^M, & M\geq 0 \\
  \tilde f^J_M(x^3){x^-}^M, & M\leq 0 \\
\end{array} \right. 
\end{align}
with $x^+=x^1+ix^2$ on the hyperboloid.
Due to the relation with the Casimir,
the spectrum of matrix d'Alembertian $\frac 1{\kappa^2}\Box$ in (\ref{matlap}) and the classical
d'Alembertian (\ref{box}) coincide.
Including also the space of polynomial functions ${\rm Pol}(X^a)$,
this is the basis for interpreting the matrix algebra $\cA$ as
quantized algebra of functions over hyperboloid.

\subsection{Semi-classical limit} 
\label{sec:semiclass}

Now consider the
classical hyperboloid $\cM$ as a Poisson manifold equipped with the Poisson structure (\ref{Poissonbra}).
The quantization of such a Poisson manifold is defined in terms of a quantization map 
$\cQ$, which is an isomorphism 
of vector spaces from the space of functions $\cC(\cM)$ on $\cM$ to 
some (operator) algebra $\cA$
\be
\begin{array}{rcl}
 \cQ:\ \ \cC(\cM) \rightarrow \cA, \qquad \; f(x) &\mapsto& \cQ(f(x))
\end{array}
\label{quant-map} 
\ee
which is compatible with the Poisson structure $\{f,g\} = \theta^{\mu\nu} \del_\mu f \del_\nu g$, 
satisfying
\begin{align}
 \cQ(f g) - \cQ(f)\cQ(g) \,\, &\to \,\, 0 \quad\mbox{and}\label{poisson-comp-1}\\
\frac 1\theta \Big(\cQ(i\{f,g\}) - [\cQ(f),\cQ(g)]\Big) \,\, &\to \,\, 0 
\qquad \mbox{as}\quad \theta \to 0 .
\label{poisson-comp}
\end{align}
Clearly $\cQ\equiv \cQ_\theta$ depends on the Poisson structure $\theta$, and the limit
$\theta\to 0$ is understood in some appropriate way; for a more mathematical discussion
we refer  e.g.  to \cite{Bordemann:1993zv}. 
As $\cQ$ is an isomorphism of vector spaces\footnote{Sometimes one only requires $Q$ to be an 
isomorphism only on the space of functions with momenta below some UV cutoff.},
one can then define the semi-classical limit of some fuzzy wavefunction $F\in\cA$
as the inverse $f = \cQ^{-1}(F)$ of the quantization map. This 
is consistent as $\theta \to 0$, provided commutators are replaced by the appropriate Poisson brackets, 
and higher order terms in $\theta$ are neglected.

In general, there is no unique way of defining $\cQ$. However 
in the case at hand, there is a natural definition of $\cQ$, based on the decomposition of 
$\cC(\cM)$ and $\cA$ into irreducible representations of $SO(2,1)$. 
Given the corresponding orthonormal bases $\Phi^J_M$ and $\phi^J_M$ of $\cA$ resp. $\cC(\cM)$
as obtained above, we define
\begin{align}
 \cQ(\phi^J_M) = \Phi^J_M, 
\end{align}
so that $\cQ$ is an isometry for the unitary representations.
This can be extended to the polynomials ${\rm Pol}(x^a)$, corresponding to finite-dimensional
non-unitary irreducible representations of $SO(2,1)$. However these are not normalizable, 
and the normalization of $\cQ({\rm Pol}(x^a))$ must be fixed in another way. 
Since we want to interpret the matrices $X^a,\;a=1,2,3$  as a quantized embedding coordinates $x^a,\;a=1,2,3$, we define
\begin{align}
 \cQ(x^a) = X^a \ .
\end{align}
Comparing the embedding equation $x^a x_a = - R^2$  with the 
Casimir constraint $X^a X_a = \kappa^2 j(j-1)$ (\ref{Casimir-constraint}), 
we are led to impose
\be\label{limeskappa}
\kappa^2 \left(s^2+\frac{1}{4}\right)=R^2 =const ,
\ee
using  $j=1/2+is$ for the principal continuous representation $\cH_j$. 
Therefore the semi-classical limit $\kappa\to 0$ implies\footnote{this is 
also the reason why the complementary series has been rejected for $\cH_j$.} $s\to \infty$.
This is the analog of $N = \dim\cH\to\infty$ in the case of fuzzy sphere.

To establish\footnote{Our aim is to establish and clarify the required properties,
without claiming mathematical rigor.} the required properties of $\cQ$, 
we recall that all fuzzy wavefunctions can be written in the following ``normal form'' \eq{matform}
\be
\Phi^J_M=F^J_M(X^3){X^+}^M, \qquad \phi^J_M=f^J_M(x^3){x^+}^M, \qquad M \geq 0
\ee
and similarly for $M<0$.
We claim that 
\be\label{smlm}
\lim_{\kappa \to 0} F^J_M=f^J_M  \ 
\ee
as functions in one variable.
To see this, observe that in the limit $\kappa \to 0$ following relations hold
\bea\label{iden-2}
&&\lim_{\kappa \to 0}\frac{1}{\kappa}[X^\pm, F(X^3)]= \mp F^\prime(X^3)X^\pm,\\
&&\lim_{\kappa \to 0}\frac{1}{\kappa}[X^-,{X^+}^M]= 2MX^3{X^+}^{M-1} ,
\eea
as a consequence of the Lie algebra relations.
In the classical case, the corresponding relations are
\bea\label{iden-3}
&& i \{X^\pm, f(x^3)\} = \mp f^\prime(x^3)x^\pm,\\
&& i\{x^-,{x^+}^M\} = 2Mx^3{x^+}^{M-1} .
\eea
Therefore the action  of the matrix Laplacian (\ref{matlap}) on $\Phi^J_M$ in the limit $\kappa \to 0$ 
\small
\bea\label{limkapnul}
\lim_{\kappa \to 0} \frac 1{\k^2}\Box \Phi^J_M=-\left[\left({X^3}^2+R^2\right){F^{\prime\prime}}^J_M(X^3)+2(M+1)X^3{F^{\prime}}^J_M(X^3)+
M(M+1)F^J_M(X^3)\right] {X^+}^M
\eea 
\normalsize
has precisely the same form as the action  of the classical Laplacian
\bea
R^2\Box_g \phi^J_M=-\frac{1}{\kappa^2}\{x^+,\{x^-,\phi^J_M\}\}+\frac{1}{\kappa^2}\{x^3,\{x^3,\phi^J_M\}\}
+\frac{i}{\kappa} \{x^3,\phi^J_M\}=-J(J-1)\phi^J_M .
\eea
This implies \eq{smlm} up to normalization, and allows to define $\cQ$ in such a way that 
$\cQ(f^J_M(x^3){x^+}^M) \to f^J_M(X^3){X^+}^M$ as $\k \to 0$. 
In particular, this provides the appropriate definition of $\cQ$ for 
the principal continuous representation $P_S$ (which is doubly degenerate), 
as well as for the finite-dimensional polynomials which are not normalizable.

Now it is easy to see that $\cQ$ respects the algebra structure and the Poisson bracket 
in the limit $\k \to 0$. Consider the product of two 
matrix modes as above, expanded up to leading order in $\kappa$ 
\bea\label{proofdeq}
\cQ(\Phi^J_M)\cQ(\Phi^{J^\prime}_{M^\prime})&=&\Phi^J_M\Phi^{J^\prime}_{M^\prime}=F^J_M {X^+}^M F^{J^\prime}_{M^\prime}{X^+}^{M^\prime}=
 F^J_M [{X^+}^M,F^{J^\prime}_{M^\prime}]{X^+}^{M^\prime}+F^J_M F^{J^\prime}_{M^\prime}{X^+}^{M+M^\prime} \nn\\
&=&(F^J_M F^{J^\prime}_{M^\prime}-M\kappa F^J_M {F^\prime}^{J^\prime}_{M^\prime}+o(\kappa^2)){X^+}^{M+M^\prime} ,
\eea
for $M, M'\geq 0$. Then \eq{poisson-comp-1} follows immediately using  \eq{smlm}.
Subtracting the same computation with the factors reversed, \eq{poisson-comp} follows.
A similar computation applies to modes with mixed or negative $M$.
Therefore $\cQ$ is indeed  a quantization map for our Poisson structure.
Using the decomposition of $\cA$ into the above modes, analogous statements can be made for the 
de-quantization map $\cQ^{-1}$, which provides the semi-classical limit of the fuzzy wavefunctions.

Finally, consider the trace of some normalizable wavefunctions with weight $M=0$,
\begin{align}
2\pi {\rm Tr} \Phi^{J^\dagger}_0 \Phi^{J}_{0}
&= 2 \pi \sum_{m=-\infty}^\infty F^{J*}_0(\kappa m)F^J_0(\k m) 
 \ \stackrel{\k\to 0}{\to} \ 2\pi\k^{-1} \int dx^3 f^{J^*}_0(x^3)f^J_0(x^3)\nn\\
&= \int \omega\, {{\phi^*}^J_0}\phi^{J}_{0},
\label{volume-trace}
\end{align}
using the explicit form of $\cH_j = P_s$ \eq{principal}, where $\omega$ is the symplectic form \eq{volel}. 
This computation is easily generalized to show
that 
\begin{align}
 2\pi {\rm Tr}\cQ(f)\cQ(g) \ \stackrel{\k\to 0}{\to} \  \int \omega \, f g
\end{align}
 as long as the integrals are bounded. This is guaranteed for the spaces of unitary wavefunctions
 discussed above.

To summarize, in the  semiclassical limit $\sim$ 
defined as de-quantization map expanded up to leading order in $\kappa$, 
we can use the following relations 
\bea\label{semicalssicallim}
\Phi^J_M\sim \phi^J_M,\;X^a\sim x^a,\;
[F,G]\sim i\{\cQ^{-1}(F),\cQ^{-1}(G)\},\;
[X^a,\;]\sim i\{ x^a,\;\},\;2\pi {\rm Tr}\sim \int \omega
\eea
which we use in the following sections.

\section{Dynamical matrix models}
Consider now three hermitian matrices $X^a = (X^a)^\dagger \in Mat(\infty,\C)$ for $a=1,2,3$, 
which transform in the
basic 3-dimensional representation of $SO(2,1)$.
Then the most general matrix model up to order 4 which is invariant under the $SO(2,1)$ symmetry 
as well as translations $X^a \to X^a + c^a \one$
has the form
\begin{align}
S[X] &=  -\frac{2\pi}{g_{YM}^{2}} {\rm Tr} \Big( [X^a,X^b][X_{a},X_{b}]  + i g_3 f_{abc} [X^a,X^b]X^c  \Big) 
\label{V-general-2}
\end{align}
for suitable constants, where embedding indices are raised and lowered with $\eta_{ab}$.
The matrices $X^a$ are understood to have dimension length, and accordingly $[g_{YM}] = L^2$.
This model is invariant under $SO(2,1)$ rotations, translations 
as well as gauge transformations $X^a \to U X^a U^{-1}$ 
for unitary operators $U$.
The equations of motion  are obtained as
\begin{align}
 4 \Box X^a  &=    3 i g_3 \, f^a_{\ bc} [X^b,X^c] \ ,   \nn\\[1ex] 
\Box &\equiv [X^a,[X_a,.]] .
\label{eom-MM-const} 
\end{align}
Now consider the ansatz\footnote{Of course the matrix model action is divergent on this background, however
this is not a problem.
We only need to require that the perturbations lead to a finite variation of the action. 
This could be taken into account by subtracting certain background terms, which we do not write down
for brevity.}
\begin{align}
 X^a &= \kappa K^a, \qquad a = 1,2,3 
\end{align}
in terms of rescaled generators of a unitary irreducible representation of $SO(2,1)$. Then
\begin{align}
 [X^a,X^b] &= i \kappa \, f^{ab}_{\ \, c}\,  X^c  ,  \nn\\
X^a X_a &= \kappa^2\, C|_{\cH} \ 
= - \kappa^2(s^2 + \frac 14)  \one_{\cH} = - R^2  \one_{\cH} \ , \nn\\
\Box X^a &=  \kappa^2\, C|_{ad} X^a = 2 \kappa^2 \, X^a  
\label{AdS-fuzzy}
\end{align}
where $C$ is the quadratic Casimir of $SO(2,1)$, and $\kappa,R$ are positive numbers. 
As discussed before, we take $\cH = \cH_j$ to be the principal continuous series representation, 
so that $X^a \in End(\cH)$ and 
\begin{align}
 \frac{R^2}{\kappa^2} = (s^2 + \frac 14) = - C|_{\cH} \ .
 \label{r-R-relation}
\end{align}
Thus the equations of motion \eq{eom-MM-const} are solved by this ansatz provided
\begin{align}
 4 \kappa^2  + 3 \kappa g_3   &= 0 .
\label{eom-coeff}
\end{align}
This is a quadratic equation in $\kappa$ which we assume to have  a positive solution.

Let us discuss the geometry of the fuzzy brane solutions in the matrix model in the semi-classical limit,
following  \cite{Steinacker:2008ri}.
Recall that the matrices $X^a$ are interpreted as quantized Cartesian embedding functions
\begin{align}
 X^a \sim x^a: \quad \cM \hookrightarrow \R^3 \ .
\end{align}
The induced metric on $\cM$ is given  by
\begin{align}
 g_{\mu\nu} = \del_\mu x^a \del_\nu x_a \ .
\end{align}
For the hyperboloid solutions under consideration $x^a x_a = -R^2$ holds, so that 
the induced metric  is that of $AdS^2$. However, 
we are interested in the effective metric which governs physical fields in the matrix model.
To identify the effective metric in the semi-classical limit, we note that the kinetic term 
(with two derivatives)
for e.g. scalar fields $\Phi$ in the matrix model\footnote{For example, the radial components of the non-abelian 
fluctuations  on a stack of coincident branes realize such scalar fields. 
A detailed analysis of general abelian perturbations will be given below.}  arises from 
an action of the form
\begin{align}
 S[\phi] &= -\frac{2\pi}{2 g_{YM}^2}\Tr\, [X_a,\Phi][X^a,\Phi] 
\sim \frac{1}{2g_{YM}^2} \int \omega\, \th^{\mu\mu'}\th^{\nu\nu'}g_{\mu'\nu'}\,\del_\mu\phi \del_\nu \phi \nn\\
 &= -\frac{1}{2g_{YM}} \int d^{2} x\sqrt{| G|} \,e^{-\sigma/2}  G^{\mu\nu} \del_\mu\phi \del_\nu \phi 
  = -\frac{1}{2} \int d^{2} x\sqrt{| G|} \,e^{-\sigma/2} G^{\mu\nu} \del_\mu\varphi \del_\nu \varphi ,
 \label{action-scalar-geom}
\end{align}
using the semi-classical correspondence rules \eq{semicalssicallim}.
Here the scalar fields are made dimensionless via  $\Phi \sim \phi = g_{YM}^{1/2}\,\varphi$, and
\begin{align}
 \omega &= \frac 12 \theta^{-1}_{\mu\nu}dx^\mu dx^\nu , \nn\\
  G^{\mu\nu} &= - g_{YM}^{-2}\,\th^{\mu\mu'}\th^{\nu\nu'}g_{\mu'\nu'}\,
   =  e^{-\sigma} g^{\mu\nu} \ ,  \nn\\
 e^{-\sigma} &=g_{YM}^{-2}|\det\theta^{\mu\nu}||\det g_{\mu\nu}|  .
\label{eff-metric}
\end{align}
Therefore  the effective metric is given by  $G^{\mu\nu}$.
Note the  explicit minus in the definition of $G^{\mu\nu}$, which is in contrast to 
the higher-dimensional case discussed in \cite{Steinacker:2008ri}. 
The correct sign is dictated by the action 
\eq{V-general-2} resp. \eq{action-scalar-geom}, which must have the form $S = \int dt (T - V)$. 
For the action \eq{V-general-2} it means that the effective metric is indeed that of $AdS^2$, while
fuzzy $dS^2$ can be obtained by changing the overall sign of the action.
This choice of signs is possible only in the case of signature $(-+)$ in 2 dimensions.
Note also that for 2-dimensional branes, the conformal factor of the effective metric is 
not fixed by the above scalar field action, due to the Weyl symmetry $G^{\mu\nu} \to e^\a G^{\mu\nu}$. 
Here we choose \eq{eff-metric} for simplicity; our main goal is to illustrate how such an  effective
metric responds to matter perturbations in the present matrix model.

The relation $G\sim  g$ is particular for 2 dimensions, and can be 
seen in coordinates where $g_{\mu\nu} = {\rm diag}(-1,1)$ at some given point $p\in \cM^2$.
Consider the point $p_N = (R,0,0)$ in the homogeneous $AdS^2$ space. 
Its tangent space is parallel to the $(x^2x^3)$ plane, so that 
we can use $x^\mu = (x^2, x^3)$ as local coordinates. 
In these ``normal embedding'' coordinates we have
$g_{\mu\nu} = {\rm diag}(-1,1)$ at $p_N$, and
$\theta^{23} = \{x^2,x^3\} =  \kappa f^{23}_{\ \, 1}\, R = \k R$. On the other hand
$\theta^{\mu\nu} = \{x^\mu,x^\nu\} =  g_{YM} e^{-\sigma/2} \epsilon^{\mu\nu}$ 
using \eq{eff-metric}, and we obtain
\begin{align}
e^{-\sigma/2} &= | g_{YM}^{-1} \,  x^1|  = g_{YM}^{-1} \kappa R  . 
\label{G-g-relation-explicit}
\end{align}
We note that 
the matrix Laplace operator \eq{AdS-fuzzy} for the unperturbed hyperboloid background
is related to the geometric Laplace operator
in the semi-classical limit\footnote{Although such a relation holds very generally in the 
higher-dimensional case  \cite{Steinacker:2008ri}, 
it is restricted to  $e^\s = const$ in 2 dimensions; for a general formula in 2 dimensions see
 \cite{arnlind}. Here we need the Laplacian only for the unperturbed backgrounds, where \eq{laplace-semiclass} is sufficient.}
\begin{align}
 \Box \Phi \sim -\{x^a,\{x_a,\phi\}\} =
 g_{YM}^2 \frac{1}{\sqrt{G}}\del_\mu(\sqrt{G} G^{\mu\nu}\del_\nu \phi) = g_{YM}^2 \Box_G \phi \ .
 \label{laplace-semiclass}
\end{align}
Finally, it is easy to add fermions the matrix model, via the action
\begin{align}
S[\psi] =  \obar\Psi\, \Gamma_a[X^a,\Psi] + m_\psi \obar\Psi \Psi \ .
\end{align}
Here 
\be
\Psi = \begin{pmatrix}
              \psi_1 \\ \psi_2
             \end{pmatrix}, \qquad \psi_\a \in \cA
\ee
is a 2-component spinors of $SO(2,1)$, and
$\Gamma_a$ satisfy the Clifford algebra of $SO(2,1)$,
\be
\Gamma_a\Gamma_b + \Gamma_b\Gamma_a = 2 \eta_{ab} \ .
\ee

\section{Fluctuating $AdS^2$ and  gravity}

We introduce some useful geometrical structures which apply to general $\cM^2\subset \R^3$.
The  ``translational currents''
\begin{align} 
 J^{a}_{\mu} &= \del_\mu x^a 
 \label{J-def}
\end{align} 
span the tangent space of $\cM^2 \subset \R^3$, while 
\begin{align} 
 K_{\mu\nu}^a &= \nabla[g]_\mu J^a_{\nu} =  K_{\nu\mu}^a 
\end{align} 
characterizes the extrinsic curvature and is normal to the brane with respect to the embedding metric,
\begin{align}
  J^{a}_{\mu} K_{a\nu\eta} = 0 \ .
\end{align}
In particular, 
\begin{align}
 K^a = K^a_{\mu\nu} G^{\mu\nu} = \Box_G x^a
\end{align}
is a normal vector\footnote{In general, this holds for
$\Box_g$ rather than $\Box_G$, but in the 2-dimensional case both statements are true.}
 to $\cM^2\subset\R^3$.
For the present $AdS^2$ solution, one can easily compute
the  curvature of the connection $\nabla[G] = \nabla[g] \equiv \nabla$, 
\begin{align}
 K_{\mu\nu}^a &=  R^{-2} g_{\mu\nu} x^a = \frac 12 G_{\mu\nu} K^a, \nn\\
 K^a &= 2 e^{-\s}R^{-2} x^a = - \Ric[G]\,  x^a \
 \label{K-explicit}
 \end{align}
 This is consistent with 
 $\Box_g x^a =  \frac {2}{R^2} x^a$ on $AdS^2$.
 The Riemann curvature tensor can be obtained e.g. from the Gauss-Codazzi theorem, 
 and is given by
 \begin{align}
{{\rm R}_{\mu\nu\eta}}^\r &= -R^{-2} (g_{\mu\eta} \d_{\nu}^{\r} - g_{\nu\eta}\d_\mu^\r) , \nn\\
{\rm Ric}_{\mu\nu} &= \frac 12 \Ric[g]\, \, g_{\mu\nu} = \frac 12 \Ric[G]\, \, G_{\mu\nu}, \nn\\
  {\rm Ric}[g] &= - 2R^{-2}, \qquad {\rm Ric}[G] = -2 R^{-2} e^{-\s} 
\label{Ricci-explicit}
\end{align}
using \eq{G-g-relation-explicit}, and recalling $x^a x_a = - R^2$.
Using the above relations along with \eq{AdS-fuzzy}, 
the embedding functions $x^a$ satisfy 
\begin{align}
 (\Box_G +\Ric[G]) x^a = 0 \ .
\label{Box-x}
\end{align}
Now consider small fluctuations around the solutions $\bar X^a$ of the above matrix model, parametrized as
\be
X^a = \bar X^a + \cA^a(\bar X) \ .
\ee
These fluctuations can be interpreted in different ways. First, one can decompose the $\cA^a$
into tangential  and one radial components, 
analogous to the well-known case of 
the fuzzy sphere \cite{Steinacker:2003sd}. 
Then the radial component can be interpreted as scalar field on 
$\cM^2$, and the tangential components in terms of (noncommutative) gauge fields.
This interpretation is certainly appropriate for the non-abelian components, which arise
on a stack of $n$ coinciding such branes. However since the trace- $U(1)$ components 
change the effective metric $G^{\mu\nu}$ on $\cM^2$, it is more natural to interpret them in 
terms of geometrical or gravitational degrees of freedom; 
note that there is no charged object under this $U(1)$.
In this section we elaborate 
some aspects of the resulting 2-dimensional effective or 
''emergent`` gravity\footnote{The word ''emergent`` indicates
 that the metric arises from other, more fundamental degrees of freedom.}.

In the semi-classical limit, the matrix model  action expanded to second order in $\cA^a$ around the basic $AdS^2$ solution is given by
\begin{align}
 S[X] &\sim \frac 2{g_{YM}^2}\int\omega\, \Big(\{x^a,\cA^b\}\{x_a,\cA_b\} +  \{x^a,\cA^b\}\{\cA_a,x_b\} 
 +  (\kappa + \frac{3g_3}2 ) f_{abc} \{\cA^a,\cA^b\} x^c \Big) \nn\\
 &= \frac 2{g_{YM}^2}\int\omega \Big(g_{YM}^2\cA_b \Box_G \cA^b - f^2 
  + (2 \kappa + \frac{3g_3}2) f_{abc} x^c \{\cA^a,\cA^b\}  \Big) \nn\\
 &= \frac 2{g_{YM}^2}\int\omega \Big(g_{YM}^2\cA_b \Box_G \cA^b - f^2 \Big)
\label{eff-S-expand}
\end{align}
dropping the linear as well as the $f_{abc}$  term which vanish due to the equations of motion 
\eq{eom-coeff}, and using 
\begin{align}
 \int \{x^a,\cA^b\}\{\cA_a,x_b\} &= - \int \cA^b\{x^a,\{\cA_a,x_b\}\} 
   =  \int \cA^b(\{\cA_a,\{x_b,x_a\}\} + \{x_b,\{x_a,\cA_a\}\}\} \nn\\
  &= \int (\kappa f_{abc} \{\cA^a,\cA^b\} x^c + \{\cA^b,x_b\},\{x_a,\cA^a\}) \ .
\end{align}
Here
\begin{align}
  f = \{\cA^a,x_a\} \   
 \label{gauge-fixing-function}
\end{align}
can be viewed as gauge fixing function, since it
transforms as 
\begin{align}
 f \to f + \{x_a,\{x^a,\L\}\} = f - g_{YM}^2\Box_G \L
\end{align}
under gauge transformations. We can thus choose the gauge such that $f=0$.

We want to understand how the geometry is influenced by matter.
We assume that all fields on $\cM^2$ couple to the effective metric\footnote{We ignore possibly different 
conformal factors for different types of matter, for the sake of illustrating the mechanism 
in a simple toy model.}
$G_{\mu\nu}$, so that the metric perturbations 
couple to matter via the energy-momentum tensor.
The linearized metric fluctuation is given by
\begin{align}
 \d_\cA g_{\mu\nu}  &= J^a_{\mu} \del_{\nu} \cA_a +  J^a_{\nu} \del_{\mu} \cA_a \nn\\
   &= \nabla_\mu \cA_\nu +  \nabla_\nu \cA_\mu - 2 K_{\mu\nu}^a \cA_a 
 \label{metric-fluct-g}
\end{align}
where we decompose the perturbations into tangential and transversal ones
\begin{align}
\cA^\perp = K^a \cA_a,   \qquad  \cA_\mu = J_\mu^a \cA_a \ .
\label{A-separation}
\end{align}
Using  $2K_{\mu\nu}^a = e^{\s} g_{\mu\nu} K^a$ \eq{K-explicit}, the perturbation of the effective metric 
in Darboux coordinates can be written as
\begin{align}
  \d_\cA G^{\mu\nu } &= -g_{YM}^{-2}\theta^{\mu\mu'}\theta^{\nu \nu '}  \d_\cA g_{\mu\nu} \nn\\
  &= -g_{YM}^{-2}\theta^{\mu\mu'}\theta^{\nu \nu '} ( \nabla_\mu \cA_\nu +  \nabla_\nu \cA_\mu)
  - e^{\s} G^{\mu\nu} \, K^a\cA_a \ .
 \label{metric-fluct}
\end{align}
Therefore
\begin{align}
 \delta_\cA S_{M} &= -\frac 12\int d^2 x \sqrt{G}\, T_{\mu\nu} \delta_\cA G^{\mu\nu} 
  = \int d^2 x \sqrt{G}  \,( \frac{e^\s}2\, T K^a  -  \nabla_\mu \tilde T^{\mu\nu}\, J^a_{\nu}) \cA_a 
\label{e-m-coupling}
\end{align}
noting that $\nabla \theta^{\mu\nu} = 0$, where
$T = T_{\mu\nu} G^{\mu\nu}$, and we  define
\begin{align}
 \tilde T^{\mu\nu}=  g_{YM}^{-2} \theta^{\nu\nu'} \theta^{\mu\mu'}T_{\mu'\nu'}
\end{align}
for convenience.
Thus the normal component $\cA^\perp$ couples to the trace of the energy-momentum tensor,
while the tangential components couple to its derivative. 
This illustrates the observation \cite{Steinacker:2012ra}
that a non-derivative coupling of the embedding perturbations to the energy-momentum tensor 
arises on branes with extrinsic curvature.
Using  the on-shell condition \eq{eom-coeff} for the background and 
\begin{align} 
\sqrt{| G_{\mu\nu}||\theta^{\mu\nu}|} = g_{YM}e^{\s/2} ,
\label{G-theta-scale}
\end{align} 
we obtain the semi-classical equations of motion
\begin{align}
 \Box_G \cA^a \ &= \
 \frac 18 g_{YM} e^{\s/2} \big(- e^\s\, T K^a + 2 J^a_\mu  \nabla_\nu \tilde T^{\mu\nu}\big) \ .
 \label{eom-matter}
\end{align} 
Note that the normal component couples to $T$ via the extrinsic curvature. This is the crucial ingredient for 
gravity, as we will see below.

\subsection{Curvature perturbations and gravity}

Now we can obtain the curvature perturbations induced by matter.
Since in 2 dimensions ${\Ric}_{\mu\nu}[G] = \frac 12 G_{\mu\nu} \Ric[G] $ 
where $\Ric[G]$ is the Ricci scalar, we will restrict ourselves to study the linearized perturbations of
$\Ric[G]$. This can be computed using
\begin{align}
   \d R_{\mu\nu}[G] &= -\frac 12 \nabla_\mu\del_\nu (G^{\rho\eta}\d G_{\rho\eta}) 
   - \frac 12 \Box_G \d G_{\mu\nu} + \nabla_{(\mu}\nabla^\eta \d G_{\nu)\eta} ,
 \end{align}
which implies
\begin{align}
 \d_\cA \Ric[G] &= (\Box_G + \frac 12 \Ric[G]) (G_{\mu\nu} \d_\cA G^{\mu\nu})
  -\nabla_\mu\nabla_\nu \d_\cA G^{\mu\nu} \ .
\end{align}
The perturbation of the effective metric in Darboux coordinates
can be written as follows (cf. \eq{metric-fluct})
\begin{align}
 \d_\cA G^{\mu\nu} =  - g_{YM}^{-2} \theta^{\mu\mu'}\theta^{\nu \nu '}
(J^a_{\mu'} \del_{\nu '} \cA_a + J^a_{\nu'} \del_{\mu '} \cA_a)
\end{align}
using $\d \theta^{\mu\nu} = 0$.
After some computations given in the appendix, the corresponding perturbation of the Ricci tensor is obtained as 
\begin{align}
 \d_\cA \Ric[G] &=
\frac 12 g_{YM} e^{\s/2} \Big(R^{-2}  \ T 
 + \nabla_\mu \nabla_\nu \tilde T^{\mu\nu}  \Big) \nn\\
 &= 8 \pi G_N \,\big(T + R^2\nabla_\mu \nabla_\nu \tilde T^{\mu\nu} \big) , \nn\\[1ex]
  8 \pi G_N &=  e^{\s/2}\, \frac{g_{YM}}{2R^2} = \frac{\kappa e^\s}{2R}  \ .
  \label{gravity-linear}
\end{align}
This can be seen as linearization of the following gravity model 
\begin{align}
 \Ric[G] - \L &= 8 \pi G_N T \ \ + \cO(\del\del T) \ , \nn\\[1ex]
\L &= -2e^{-\s}R^{-2} \ ,
\end{align}
which is reasonable and non-trivial in 2 dimensions 
\cite{Brown:1986nm} (dropping the $\cO(\del\del T)$ terms), unlike
general relativity which does not allow any coupling to matter.
Note that the derivative term is of order 
\begin{align}
  \nabla\nabla\tilde T \sim e^{\s} \nabla\nabla T
\end{align}
using \eq{G-theta-scale},
and can be neglected provided $e^{\s} \ll 1$, which is compatible with $G_N \ll 1$.
Although we focused on the $AdS^2$ background, the result should equally apply to the $dS^2$ background,
which is obtained by changing the sign of the matrix model action.

We emphasize again that no specific gravity action was assumed or induced, we have simply 
elaborated the matrix model dynamics from a geometrical point of view.
The crucial coupling to $T_{\mu\nu}$ arises due to the extrinsic curvature of the brane
encoded in $\nabla_\mu J^{a}_\nu  = K_{\mu\nu}^a$, as pointed out in  \cite{Steinacker:2012ra};
this is already seen in \eq{eom-matter}.
Also, it is gratifying (and not evident) that the Newton constant turns out to be positive.
The  mechanism is basically the 
same as the ''gravity bag`` mechanism discussed in  \cite{Steinacker:2009mp}.
Its 4-dimensional version is clearly more complicated
and currently under investigation, however at least certain aspects 
of the mechanism generalize \cite{Steinacker:2012ra}.

However, since the gravitational coupling is dynamical itself, the above linearized treatment 
of the coupling is justified only as 
long as the perturbations of the radial $K_{\mu\nu}^a$ is negligible, i.e.
\begin{align}
 \d K_{\mu\nu}^a  \ll K_{\mu\nu}^a \ .
\end{align}
 For the  $AdS^2$ backgrounds under consideration, this 
implies that the intrinsic curvature perturbation is smaller than the background constant 
curvature. This is clearly inadequate for physical gravity, however 
the basic mechanism  should extend beyond this regime 
for backgrounds where the extrinsic curvature dominates 
the intrinsic one, such as cylinders or generalizations.

 \subsection{Induced metric curvature}
 
 It is  instructive to compute also the Ricci tensor for the induced metric $g_{\mu\nu}$.
 Recall the decomposition of $\cA_a$ into tangential and normal components \eq{A-separation}.
 We have
  \begin{align}
 \d_\cA \Ric[g] &= -(\Box_g + \frac 12 \Ric[g]) (g^{\mu\nu} \d_\cA g_{\mu\nu}) 
 +\nabla^\mu\nabla^\nu \d_\cA g_{\mu\nu} \ .
\end{align}
Writing the metric perturbation as
 \begin{align}
  \d_\cA g_{\mu\nu} &=  \nabla_\mu \cA_\nu +  \nabla_\nu \cA_\mu - 2 R^{-2} g_{\mu\nu} x^a \cA_a \ ,
 \end{align}
one finds
\begin{align}
 \nabla^\mu\nabla^\nu \d_\cA g_{\mu\nu} 
 &= \nabla^\mu\nabla^\nu( \nabla_\mu \cA_\nu +  \nabla_\nu \cA_\mu - 2 R^{-2} g_{\mu\nu} x^a \cA_a) \nn\\
 &= 2\Box_g (\nabla^\nu\cA_\nu) + \Ric[g] (\nabla^\nu\cA_\nu) - 2 R^{-2} \Box_g(x^a \cA_a)\ ,
\end{align}
noting that $\Ric_{\mu\nu}[g] = \frac 12 \Ric[g] g_{\mu\nu}$ and the identity \eq{Box-del-A}.
Therefore
\begin{align}
 \d_\cA \Ric[g] &= -(2\Box_g + \Ric[g]) ( \nabla^\nu\cA_\nu - 2 R^{-2} x^a \cA_a) \nn\\
 &  + 2\Box_g (\nabla^\nu\cA_\nu) + \Ric[g] (\nabla^\nu\cA_\nu) - 2 R^{-2} \Box_g(x^a \cA_a) \nn\\
 &=  2 R^{-2} (\Box_g + \Ric[g])(x^a \cA_a) \ .
\end{align}
As a consistency check, we 
note that the tangential variations $\cA_\mu$ drop out, since they correspond to a diffeomorphism.
Since $g =  e^\s G$, 
this is related to $\d_\cA \Ric[G]$ up to conformal rescaling contributions.

\subsection{Gauge theory point of view}

In this final section, we disentangle and  essentially solve the model  using the gauge theory point of view.
Recall the  decomposition \eq{A-separation} of $\cA^a$ into normal and tangential components.
For the normal perturbations $\cA^\perp$, we can use the identity
\begin{align}
 \Box_G \cA^\perp &= -\Ric[G]\,(K^a A_a) + 2 \, \nabla_\mu K^a \del^\mu \cA_a + K^a \Box_G A_a \nn\\
&= \Ric[G]\, \cA^\perp - 2 \Ric\,(\nabla^\mu\cA_\mu) + K^a \Box_G A_a \ ,
\label{wave-radial-id} 
\end{align} 
so that  using the equation of motion \eq{eom-KBoxA} gives
\begin{align}
 \Box_G \cA^\perp &= \Ric\, \cA^\perp - 2 \Ric\,(\nabla^\mu\cA_\mu) 
 + \frac 12 \k\, R^{-1}  \ T \ .
\label{wave-eq-radial} 
\end{align} 
Similarly, consider the  divergence of the tangential perturbations 
\begin{align}
 \nabla^\mu \cA_\mu =  \nabla^\mu(J_\mu^a \cA_a) = K^a \cA_a + J_\mu^a\nabla^\mu\cA_a \ .
\end{align}
The tangential components of the equation of motion give
 \begin{align}
 J^a_\mu \Box_G \cA_a \ &= \
  \frac 14 g_{YM} e^{\s/2} g_{\eta\mu}  \nabla_\nu \tilde T^{\eta\nu} \ ,
 \label{eom-matter-tang}
\end{align} 
so that 
\begin{align}
 \Box_G \cA_\mu &= 2 \nabla_\r  J_\mu^a \del^\r \cA_a + \Box_G J^a_\mu \cA_a  
  +\frac 14 g_{YM} e^{\s/2} g_{\eta\mu}  \nabla_\nu \tilde T^{\eta\nu}  \nn\\
  &=  2 K_{\r\mu}^a \del^\r \cA_a - \frac 12 {\Ric[G]}\, \cA_\mu 
 +\frac 14 g_{YM} e^{-\s/2} G_{\eta\mu}  \nabla_\nu \tilde T^{\eta\nu}  \ 
\end{align}
using $\nabla^\mu K_{\mu\nu}^a = \Box_G J_\mu^a = - \frac 12 {\Ric[G]} J^{a}_\mu$
and $K_{\r\mu}^a = \frac 12 G_{\r\mu} K^a$. This gives 
\begin{align}
 \nabla^\mu\Box_G \cA_\mu
  &=  -\Ric[G]\, J^a_\r \del^\r \cA_a + K^a \Box_G \cA_a
 - \frac 12 {\Ric[G]}\,\nabla^\mu \cA_\mu 
 +  \frac 14 g_{YM} e^{-\s/2}  \nabla_\mu \nabla_\nu \tilde T^{\eta\nu} \nn\\
 &=  \Ric[G]\, K^a \cA_a - \frac 32 \Ric[G]\,\nabla^\mu \cA_\mu
 + \frac 14 \kappa R \big(2 R^{-2}  \ T + \nabla_\mu \nabla_\nu \tilde T^{\mu\nu} \big) \ .
\end{align}
 Together with 
\begin{align}
 \Box_G(\nabla^\mu \cA_\mu) 
 &= -\frac 12 {\rm Ric[G]} \,\nabla^\mu \cA_\mu + \nabla^\mu \Box_G \cA_\mu \ ,
\label{Box-del-A}
\end{align}
it follows that the scalar field $\nabla^\mu\cA_\mu$ satisfies the  wave equation 
\begin{align}
 \Box_G (\nabla^\mu\cA_\mu) &= -2 \Ric[G]\, (\nabla^\mu\cA_\mu) + \Ric[G]\, \cA^\perp
 + \frac 14 \kappa R \big(2 R^{-2}  \ T + \nabla_\mu \nabla_\nu \tilde T^{\mu\nu} \big) \ .
\end{align}
Together with \eq{wave-eq-radial} we 
 we obtain the following ''almost-decoupled`` wave equations
\begin{align}
  \Box_G \chi &= \frac{\kappa R}{4}  \nabla_\mu \nabla_\nu \tilde T^{\mu\nu}  \ , \nn\\
  (\Box_G + \Ric[G]) \cA^\perp &= - 2 \Ric[G]\,\chi + \frac 12 \k\, R^{-1}  \ T \ ,
  \label{full-wave-eq}
\end{align}
where
 \begin{align}
 \chi:= \nabla^\mu \cA_\mu - \cA^\perp  = J_\mu^a \del^\mu \cA_a \ .
 \end{align}
The second is a scalar wave equation for $\cA^\perp$, and $\chi$ can be seen as part of its source,
determined by the first equation.
For distances below the ''cosmological`` scales, the 
mass term can be neglected, leading to massless wave equations with source determined by $T_{\mu\nu}$
as above.

A remark on the relation with the noncommutative gauge theory point of view is in order.
The usual gauge fields $A_\mu$ in the gauge theory interpretation are related to 
our tangential perturbations as
\begin{align}
 \theta^{\mu\nu}A_\nu = \eta^{\mu\nu}\cA_\nu \ ,
\end{align}
since $J_{\mu a} = \eta_{\mu a}$ if $x^a$ for $a = 0,1$ are normal embedding coordinates,
cf. \eq{G-g-relation-explicit}.
Thus
\begin{align}
 \del^\mu \cA_\mu \sim  \theta^{\mu\nu}\del_\mu A_\nu = \frac 12  \theta^{\mu\nu} F_{\mu\nu}
\end{align}
up to some constant. This is  gauge invariant (more precisely it
transforms as a  scalar field under noncommutative gauge transformations i.e.  symplectomorphisms),
and encodes the only physical degree of freedom in 2D gauge theory.
Similarly, $\cA^\perp$ can be interpreted as noncommutative scalar field in the 
noncommutative gauge theory.
Therefore $ \del^\mu \cA_\mu$ and $\cA^\perp$ completely capture the physics of the 
system, which is described by \eq{full-wave-eq} 
at the semi-classical (Poisson) level.
It is also worth pointing out that the radial and tangential perturbations mix as 
observed in  \cite{Steinacker:2012ra}, but we were able to disentangle them 
 in the 2-dimensional case.

\section{Conclusion}

We studied the fuzzy version of 2-dimensional de Sitter and Anti-de Sitter space, and some of the associated physics. 
The quantization map is discussed in detail, and we obtained explicit formulae for the 
functions on the fuzzy hyperboloid corresponding to unitary irreducible representations of $SO(2,1)$.
This should provide the basis for further work on the associated non-commutative
field theory on a curved space-time with Minkowski signature. 
Moreover, we consider a matrix model which admits fuzzy $(A)dS^2$ as solution, and study the resulting dynamics
of the geometry. This allows to study the general ideas of emergent geometry in matrix models
on a simple curved background with Minkowski signature. 
Although the model is modified 
as compared with the IKKT model by adding a cubic term, it is  an interesting toy model
which allows to essentially solve the resulting dynamics. We find that the transversal brane perturbations
indeed couple to the energy-momentum tensor as emphasized in \cite{Steinacker:2012ra}, and we also find
a mixing between  tangential and transversal perturbations in the gauge theory point of view.
The brane dynamics leads to a reasonable linearized gravity theory, related to  Henneaux -- Teitelboim
gravity in 2 dimensions. 
It is remarkable that this happens through the bare matrix model action, without adding any gravity terms 
and without invoking any quantum effects. The mechanism does not require a strong-coupling regime.
Even though the present toy model is not of direct physical relevance, it is nevertheless  useful 
to clarify the dynamics of the branes and their geometry, as a step towards higher-dimensional more 
physical matrix models such as the IKKT model.

It would also be interesting to study a finite-dimensional realization of the matrix model numerically,
following \cite{Kim:2011ts}. 
This might serve as a toy model and testing ground for the case of Minkowski signature, as a step 
towards the higher-dimensional case.

\paragraph{Acknowledgments.}

The work of H.S. is supported by the Austrian Fonds f\"ur Wissenschaft und Forschung under grant P24713,
and the work of D.J. was supported by the PostDoc program of the 
Croatian Science Fundation.
This collaboration was also supported by the Austrian-Croatian WTZ project HR 22/2012 of the OEAD.

\section{Appendix: Derivation of the linearized gravity equations}

We  note the following identities
\begin{align}
 J^a_{\mu} J_{a\nu} &= g_{\mu\nu}, \qquad  J^a_{\mu} K_{a \nu\eta} = 0
\end{align}
as well as
\begin{align}
  \nabla^\mu J_\mu^a &= K^a \ , \nn\\
 \Box_G J_\mu^a &= \del_\mu \Box_G x^a + {\rm Ric}_{\mu\nu} J^{a\nu} \ 
 = - \frac 12 \Ric[G]\,J^{a}_\mu
\label{J-id-2}
\end{align}
which follows from \eq{Box-x}.
Then
\begin{align}
  g_{YM}^{-2}\theta^{\mu\mu'}\theta^{\nu \nu '}\nabla_\mu\nabla_\nu(J^a_{\nu'} \del_{\mu'} \cA_a) 
  &= g_{YM}^{-2}\theta^{\mu\mu'}\theta^{\nu \nu '}\nabla_\mu(K_{\nu\nu'}^a\del_{\mu'} \cA_a 
     +  J^a_{\nu'} \nabla_\nu\del_{\mu'} \cA_a) \nn\\
  &= g_{YM}^{-2}\theta^{\mu\mu'}\theta^{\nu \nu '}(K^a_{\mu\nu'} \nabla_\nu\del_{\mu'} \cA_a
    + J^a_{\nu'}\nabla_\mu \del_{\mu'} \nabla_\nu\cA_a ) \nn\\
  &=  \frac {e^\s}{2} K^a\Box_G \cA_a
  +\frac 12 g_{YM}^{-2}\theta^{\mu\mu'}\theta^{\nu \nu '} J^a_{\nu'}\, {{\rm R}_{\mu\mu';\nu}}^{\r} \del_{\r} \cA_a \nn\\ 
  &=  \frac {e^\s}{2} K^a\Box_G \cA_a
  -\frac 12  g_{YM}^{-2}R^{-2}\theta^{\mu\mu'}\theta^{\nu \nu '} J^a_{\nu'} 
  (g_{\mu\nu} \d_{\mu'}^{\r} - g_{\mu'\nu}\d_{\mu}^{\r}) \del_{\r} \cA_a \nn\\ 
 &=  \frac {e^\s}{2} K^a\Box_G \cA_a + R^{-2} J^a_{\nu}\del^{\nu} \cA_a  
\end{align}
using \eq{Ricci-explicit}, and noting that $\nabla[g] = \nabla[G]$ here.
Similarly,  we obtain using \eq{J-def}
\begin{align}
  g_{YM}^{-2} \theta^{\mu\mu'}\theta^{\nu \nu '}\nabla_\mu\nabla_\nu(J^a_{\mu'} \del_{\nu '} \cA_a) 
  &= g_{YM}^{-2}\theta^{\mu\mu'}\theta^{\nu \nu '}\nabla_\mu(K_{\nu\mu'}^a \del_{\nu '} \cA_a +  J^a_{\mu'} \nabla_\nu\del_{\nu '} \cA_a) \nn\\
  &=  \frac {e^\s}{2} (K^a\Box_G \cA_a + \del_\mu K^a \del^\mu \cA_a) \nn\\
 &=  \frac {e^\s}{2} K^a\Box_G \cA_a + R^{-2} J^a_{\nu}\del^{\nu} \cA_a  .
\end{align}
Therefore
\begin{align}
 e^{-\s}\nabla_\mu\nabla_\nu \d_\cA G^{\mu\nu} = - K^a\Box_G \cA_a + \Ric[G]\, J^a_{\nu}\del^{\nu} \cA_a \ .
\end{align}
Finally, we have
\begin{align}
\frac 12 e^{-\s} G_{\mu\nu}\d G^{\mu\nu} &= e^{-\s} g^{\mu\nu} J^a_{\mu} \del_{\nu} \cA_a 
 = J^a_{\mu} \del^{\mu} \cA_a \ .
 \label{GdG}
\end{align}
Therefore  
\begin{align}
e^{-\s} \d_\cA \Ric[G] &= e^{-\s}(\Box_G + \frac 12 \Ric[G]) ( G_{\mu\nu} \d G^{\mu\nu}) - e^{-\s}\nabla_\mu\nabla_\nu \d G^{\mu\nu} \nn\\
 &= (2\Box_G + \Ric[G])(J^a_{\mu} \del^{\mu} \cA_a)
 + (K^a\Box_G \cA_a -\Ric[G]\, J^a_{\nu}\del^{\nu} \cA_a) \nn\\
&=  K^a \Box_G \cA_a + 2 \nabla^{\mu}(J^a_{\mu}\Box_G  \cA_a) \ ,
\end{align}
where we used 
\begin{align}
 \Box_G (J^a_{\mu} \del^{\mu} \cA_a) &= \Box_G J^a_{\mu} \del^{\mu} \cA_a +  J^a_{\mu}\Box_G \del^{\mu} \cA_a 
  + 2 K^a_{\mu\nu}\nabla^\mu\nabla^\nu \cA_a   \nn\\
&= -\frac 12  \Ric[G] J^a_{\mu}\del^{\mu}\cA_a + J^a_{\mu}\del^{\mu}\Box_G  \cA_a + \Ric_{\mu\nu}[G] J^a_{\mu}\del^{\nu}\cA_a
 +  K^a \Box_G \cA_a \nn\\
&= \nabla^{\mu}(J^a_{\mu}\Box_G  \cA_a)  - (\nabla^{\mu}J^a_{\mu})\Box_G  \cA_a 
 +  K^a \Box_G \cA_a \nn\\
&= \nabla^{\mu}(J^a_{\mu}\Box_G  \cA_a)
\end{align}
due to \eq{J-id-2}.
Now we can use the equations of motion \eq{eom-matter}, which give
\begin{align}
 \nabla^{\mu}(J^a_{\mu}\Box_G  \cA_a) &= \frac 14 g_{YM} e^{\s/2} \nabla^\mu g_{\mu\nu}\nabla_\r \tilde T^{\nu\r} 
= \frac 14 g_{YM} e^{-\s/2} \nabla_\mu \nabla_\nu \tilde T^{\mu\nu} \nn\\
 K^a \Box_G \cA_a &= -\frac 18 g_{YM} e^{3\s/2}\,  K^a K_a \ T
  =  \frac 12 g_{YM} e^{-\s/2}\, R^{-2}  \ T
\label{eom-KBoxA}
\end{align}
recalling that $J^a K^a = 0$, as well as
\begin{align}
  K^a K_a = - 4 e^{-2\s} R^{-2} \ .
\end{align}
Putting these together, we finally arrive at 
\begin{align}
 \d_\cA \Ric[G] &=
\frac 12 g_{YM} e^{\s/2} \Big(R^{-2}  \ T 
 + \nabla_\mu \nabla_\nu \tilde T^{\mu\nu}  \Big) \nn\\
 &= 8 \pi G_N \,\big(T + R^2\nabla_\mu \nabla_\nu \tilde T^{\mu\nu} \big)  \nn\\[1ex]
  8 \pi G_N &=  e^{\s/2}\, \frac{g_{YM}}{2R^2} = \frac{\kappa e^\s}{2R}  \ .
  \label{gravity-linear-app}
\end{align}

\end{document}